\documentclass{elsart}

\usepackage{amssymb}

\begin{document}

\begin{frontmatter}




\title{Quantum walks and \\ reversible cellular automata}
\author[Yokohama]{N. Konno\corauthref{cor}},
\corauth[cor]{Corresponding author.}
\ead{norio@mathlab.sci.ynu.ac.jp}
\author[Yokohama]{K. Mistuda},
\author[Yokohama]{T. Soshi},
\author[Yonsei]{H. J. Yoo}

\address[Yokohama]{Department of Applied Mathematics, Yokohama National University, Hodogaya-ku, Yokohama, 240-8501, Japan}
\address[Yonsei]{University College, Yonsei University, Seodaemoon-gu, Seoul, 120-749, Korea}

\begin{abstract}
We investigate a connection between a property of the distribution and a conserved quantity for the reversible cellular automaton derived from a discrete-time quantum walk in one dimension. As a corollary, we give a detailed information of the quantum walk.
\end{abstract}

\end{frontmatter}

\newcommand{\ket}[1]{|#1\rangle}
\newcommand{\bra}[1]{\langle#1|}
\newcommand{\U}{\bar{U}}
\newcommand{\braa}{\langle}
\newcommand{\kett}{\rangle}


\section{Introduction}
Recently considerable work has been done on quantum walks (QWs), which may be useful in constructing efficient quantum algorithms. Kempe \cite{Kempe}, Tregenna {\it et al.} \cite{Tregenna}, Ambainis \cite{Ambainis} have reviewed the field. For more general setting in connection with quantum cellular automata, see Meyer \cite{Meyer}. Two classes of QWs have been studied: one is a discrete-time case, the other is a continuous-time case. Here we focus on discrete-time QWs in one dimension. The QW can be considered as a quantum analog of the classical random walk. However there are some differences between them. For example, a QW starting from the origin at time $n=0$ moves away faster than the classical random walk. The standard deviation $\sigma (n)$ of the probability distribution for the QW increases with time according to $\sigma (n) \sim n$, while for the classical random walk, $\sigma (n) \sim \sqrt{n}.$ 

In this Letter we consider some properties on a one-dimensional reversible cellular automaton (RCA) derived from a QW on a line. We present necessary and sufficient conditions on the initial state for some conserved quantities of the RCA. One is the expectation of the distribution (the 1st moment), and the other is the squared norm of the distribution (the 0th moment). The former corresponds to the symmetry of distribution, because our result (see Theorem 2) implies that the distribution is symmetry for any time step if and only if its expectation becomes zero for any time step. We should note that our RCA is different from a quantum cellular automaton studied by Gr\"ossing and Zeilinger \cite{GZa,GZb}. In \cite{GZb}, they found a conservation law connects the strength of the mixing of locally interacting states and the periodicity of the global structures. Their conservation law also differs from our conserved quantities. Concerning a recent review on quantum cellular automata, see Aoun and Tarifi \cite{Aoun}, and Schumacher and Wernerfor \cite{Schumacher}, for examples. It is well known that cellular automata are models based on simple rules which upon deterministic time evolution exhibit various complex behavior \cite{Wolfram}. So our investigation on the RCA might be applicable to the analysis on the behavior. Moreover, by applying our results to the QW case, we obtain a more detailed information on QWs. 

The rest of the Letter is organized as follows. Section 2 gives the definitions of the QW and the RCA. In Section 3, we study a necessary and sufficient condition on the initial state of the symmetry of distribution (in general, non-probability distribution) of the RCA. Section 4 treats a necessary and sufficient condition on the initial state for a conserved quantity of the RCA. In the final section, we apply our results to the QW.

\section{Definitions of the QW and the RCA}
In this section first we present a definition of the QW. The time evolution of the QW is given by the following unitary matrix:
\begin{eqnarray*}
U=
\left[
\begin{array}{cc}
a & b \\
c & d
\end{array}
\right]
\end{eqnarray*}
\par\noindent
where $a,b,c,d \in {\bf C}$ and ${\bf C}$ is the set of complex numbers. The unitarity of $U$ gives $|a|^2 + |c|^2 =|b|^2 + |d|^2 =1, \> a \overline{c} + b \overline{d}=0, \> c= - \triangle \overline{b}$ and $d= \triangle \overline{a},$ where $\overline{z}$ is a complex conjugate of $z \in {\bf C}$ and $\triangle = \det U = ad - bc.$ The QW is a quantum generalization of the classical random walk with an additional degree of freedom called the chirality. The chirality takes values {\it left} and {\it right}, and means the direction of the motion of the particle. At each time step, if the particle has the left chirality, it moves one step to the left, and if it has the right chirality, it moves one step to the right. 

The evolution is defined by the following way. The unitary matrix $U$ acts on two chirality states $\ket{L}$ and $\ket{R}$: $\ket{L} \> \to \> a\ket{L} + c\ket{R}$ and $\ket{R} \> \to \> b\ket{L} + d\ket{R},$ where $L$ and $R$ refer to the right and left chirality state respectively. In fact, put $\ket{L} = {}^t [1,0]$ and $\ket{R} = {}^t [0,1]$ where $t$ means the transposed operator. So we have $U\ket{L} = a\ket{L} + c\ket{R}$ and $U\ket{R} = b\ket{L} + d\ket{R}.$ More in detail, at any time $n$, a 2-vector $\in {\bf C}^2$ at the location $k \in {\bf Z},$ where ${\bf Z}$ is the set of integers, gives the amplitude of the probability of finding the particle at $k$: the probability is given by the square of the absolute value of the vector at $k$. If $\ket{\Psi_k(n)}$ define the amplitude at time $n$ at the location $k$ where
$$
\ket{\Psi_k{(n)}} = 
\left[
\begin{array}{cc}
\Psi_{k}^{L} (n)\\
\Psi_{k}^{R} (n)   
\end{array}
\right] \in {\bf C}^2
$$
with the chirality being left (upper component) or right (lower component), then the dynamics for $\ket{\Psi_k{(n)}}$ in the QW is described by the following transformation:
\begin{eqnarray}
\ket{\Psi_k{(n+1)}} = \ket{L}\bra{L}U\ket{\Psi_{k+1}{(n)}}+\ket{R}\bra{R}U\ket{\Psi_{k-1}{(n)}}
\label{eqn:def}
\end{eqnarray}   
for any time $n \in {\bf Z}_+,$ where ${\bf Z}_+$ is the set of non-negative integers. It is known that the evolution of the QW is highly sensitive to the initial qubit state (see Refs. \cite{KonA,KonB,KonC,KonD}, for examples). Therefore we define the set of initial qubit states as follows:
\[
\Phi = \left\{ \varphi =
{}^t \left[ \Psi_0^L(0), \Psi_0^R(0) \right] \in {\bf C}^2
:
|\Psi_0^L(0)|^2 + |\Psi_0^R(0)|^2 =1
\right\}
\]   
Here we introduce some typical QWs. The simplest and well-studied QW is the Hadamard walk whose unitary matrix $U$ is defined by 
\begin{eqnarray*}
H = 
{1 \over \sqrt{2}}
\left[
\begin{array}{cc}
1 & 1 \\
1 & -1 
\end{array}
\right] 
\end{eqnarray*}
The dynamics of this walk corresponds to that of the symmetric random walk in the classical case. However, it is noted that the symmetry of the distribution for the Hadamard walk depends heavily on the initial qubit state, see Konno {\it et al.} \cite{KonA} (the Hadamard case), Konno \cite{KonB,KonC,KonD} (general case). Another example is an extension of the Hadamard walk in the following way (see Romanelli {\it et al.} \cite{Romanelli}, for instance):
\begin{eqnarray*}
H (\theta) = 
\left[
\begin{array}{cc}
\cos \theta & \sin\theta \\
\sin\theta & - \cos\theta 
\end{array}
\right] 
\end{eqnarray*}
where $\theta \in [0,2 \pi)$. In this Letter we consider mainly the case of $\theta \in [0, \pi/2)$ for the sake of simplicity. Remark that $H(\pi/4)=H$. Now  Eq. (\ref{eqn:def}) can be written as
\begin{eqnarray*}
\Psi_k^L(n+1)=a \Psi_{k+1}^L(n)+b \Psi_{k+1}^R(n)
\\
\Psi_k^R(n+1)=c \Psi_{k-1}^L(n)+d \Psi_{k-1}^R(n)
\end{eqnarray*}
Furthermore, we can apply the above equations to each other to uncouple the chirality components. Resultantly, both $\Psi_k^L(n)$ and $\Psi_k^R(n)$ satisfy the following same partial difference equation:
\begin{eqnarray}
X_k(n+2)=a X_{k+1}(n+1)+d X_{k-1}(n+1)-(ad-bc) X_{k}(n)
\label{eqn:defRCA}
\end{eqnarray}
The above argument appeared in Knight {\it et al.} \cite{KnightA,KnightB} (see also page 279 in Gudder \cite{Gudder}). In other words, Eq. (\ref{eqn:defRCA}) implies a dynamical independence of the evolution of the two chiralities $L$ and $R$. Thus there are two essentially independent walks, coupled only by the first two steps. After that the two walks can behave independently each other. In this Letter, we call the cellular automaton defined by Eq. (\ref{eqn:defRCA}) {\it reversible cellular automaton (RCA)$\>$}, since Eq. (\ref{eqn:defRCA}) implies that $X_{k}(n)$ can be also determined by $X_{k}(n+1)$ and $X_{k}(n+2)$. In the terminology of cellular automata, the state space at each lattice site is usually taken to be finite, however, continuous case can also be chosen (see pp.155-160 in \cite{Wolfram}, for instance). A special case of $a=d=1$ and $b=c=0$ is equivalent to a discretized two-dimensional wave equation derived from numerical modeling of cosmic strings, see Ref. \cite{Winitzki}.

Here we give an important relation between the QCA and the QW. If we put $(X_{0}(0), X_{-1}(1), X_{1}(1))=(\Psi_0^L(0), a\Psi_0^L(0)+b\Psi_0^R(0), 0)$ on the initial state of Eq. (\ref{eqn:defRCA}), then $X_k(n)=\Psi_k^L(n)$, that is, $X_k(n)$ represents the left chirality of the QW. Similarly, if $(X_{0}(0), X_{-1}(1), X_{1} (1) ) = (\Psi_0^R (0), 0, c \Psi_0^L (0) + d \Psi_0^R(0) )$ is chosen, then $X_k(n)=\Psi_k^R(n)$, i.e., $X_k(n)$ represents the right chirality of the QW.

In this Letter we consider a more general setting concerning the initial state. So we define the set of initial states for the RCA:
\[
\widetilde{\Phi} = \left\{ \widetilde{\varphi} =
(X_{0}(0), X_{-1}(1), X_{1}(1))\equiv (\alpha, \beta, \gamma) 
: \alpha, \beta, \gamma \in {\bf C}
\right\}
\]
To distinguish our initial state for the RCA considered here from that for the QW in our previous papers \cite{KonA,KonB,KonC,KonD}, we put ``$\sim$'' in $\Phi$ etc. From now on we consider only the RCA defined by $H(\theta)$:
\begin{eqnarray}
X_k(n+2)=\cos\theta [X_{k+1}(n+1)- X_{k-1}(n+1)] +X_{k}(n)
\label{eqn:naomi}
\end{eqnarray}
Here we define a distribution (in general, non-probability distribution) of the RCA $X(n)$ at time $n \in {\bf Z}_+$ by $\{ |X_{k}(n)|^2 : k \in {\bf Z} \}.$

McGuigan \cite{McGuigan} investigated some classes of quantum cellular automata. In his setting, Eq. (\ref{eqn:naomi}) belongs to a class of fermionic quantum cellular automata whose update equation is given by 
\begin{eqnarray*}
X_k(n+2)= f( X_{k+1}(n+1), X_{k}(n+1), X_{k-1}(n+1)) +X_{k}(n)
\label{eqn:naomisan}
\end{eqnarray*}
Our case is $f(x,y,z) = (x-z) \> \cos \theta.$ By using Eq. (\ref{eqn:naomi}), Romanelli {\it et al.} \cite{Romanelli} analyzed in detail discrete-time one-dimensional QWs by separating the quantum evolution into Markovian and interference terms.

\section{Symmetry of distribution of the RCA}
First we present the following useful lemma to prove Theorem 2. 
\begin{lem}
\label{lem:lem1}
\par\noindent
{\rm (i)} Suppose that initial state is 
\[
\widetilde{\varphi} 
= (X_{0}(0), X_{-1}(1), X_{1}(1))\equiv (\alpha, \beta, -\beta) 
\]
where $\alpha,\beta \in {\bf C}$. Then, for any $k, n \in {\bf Z}_+$, 
\begin{eqnarray}
X_k(n)=(-1)^n X_{-k}(n)
\label{eqn:yuki}
\end{eqnarray}
\par\noindent
{\rm (ii)} Suppose that initial state is 
\[
\widetilde{\varphi} 
= (X_{0}(0), X_{-1}(1), X_{1}(1))\equiv (0, \beta, e^{i \xi}\beta) 
\]
where $\beta, \xi \in {\bf R}$ and ${\bf R}$ is the set of real numbers. Then, for any $k, n \in {\bf Z}_+$, 
\begin{eqnarray}
X_k(n)=(-1)^{n+1} e^{i \xi} \> \overline{X_{-k}(n)}
\label{eqn:aki}
\end{eqnarray}
\end{lem}
\par\noindent
The above lemma directly comes from the symmetry: if $X_k(n)$ is a solution of the RCA, then so is $X^{(1)}_k(n)=(-1)^n X_{-k}(n)$, and also so is, for any $\xi$, $\> X^{(2)}_k(n)=(-1)^{n+1} e^{i \xi} \> \overline{X_{-k}(n)}.$ To state Theorem 2, we introduce the following three classes:
\begin{eqnarray}
\widetilde{\Phi}_{\bot} &=& \left\{ \widetilde{\varphi} 
\in 
\widetilde{\Phi} :
\beta + \gamma =0  \right.
\label{eqn:yama}
\\
&&
\hspace{3mm}\> \hbox{or} \>  |\beta|= |\gamma|(>0), 
\> \hbox{and}\>\> \alpha =0
\label{eqn:tano}
\\
&& 
\hspace{3mm}\> \hbox{or} \>  
|\beta|= |\gamma|(>0), \alpha \not= 0, \> 
\hbox{and}\>\> \theta_{\beta}+\theta_{\gamma}-2\theta_{\alpha} =\pi \>\>
( \hbox{mod} \> 2\pi)\left.\right\} 
\label{eqn:oroti}
\\
\widetilde{\Phi}_s &=&  \{\widetilde{\varphi} \in 
\widetilde{\Phi} : \> 
|X_{k}(n)| = |X_{-k}(n)| \>\>
\hbox{for any} \> n \in {\bf Z}_+ \> \hbox{and} \> k \in {\bf Z} 
\} 
\nonumber
\\
\widetilde{\Phi}_0 &=& \left\{ \widetilde{\varphi} \in 
\widetilde{\Phi} : \> 
\sum_{k=-\infty}^{\infty}k |X_{k}(n)|^2=0 \>\> \hbox{for any} \> n \in {\bf Z}_+\right\}
\nonumber
\end{eqnarray}
where $\theta_{z}$ is the argument of $z \in {\bf C}$ with $z \not= 0$. It is noted that if $\widetilde{\varphi} \in \widetilde{\Phi}_s$, then the distribution of $X(n)$ is symmetric for any time step $n \in {\bf Z}_+$.  
\begin{thm}
\label{thm:thm2} For the RCA defined by Eq. (\ref{eqn:naomi}), we have
\begin{eqnarray*}
\widetilde{\Phi}_{\bot} = \widetilde{\Phi}_s = \widetilde{\Phi}_0
\end{eqnarray*}
\end{thm}
\par\noindent
Proof. First we see that the definitions of $\widetilde{\Phi}_s$ and $\widetilde{\Phi}_0$ give immediately 
\begin{eqnarray}
\widetilde{\Phi}_s \subset \widetilde{\Phi}_0 
\label{eqn:tsuika}
\end{eqnarray}
\par
Next we prove  $\widetilde{\Phi}_{\bot} \subset \widetilde{\Phi}_{s}$.
The proof is divided into the following three cases. \\
Case 1. If an initial state satisfies Eq. (\ref{eqn:yama}), then we can obtain the following relation from Lemma 1 (i):
\begin{eqnarray}
|X_k(n)|=|X_{-k}(n)|
\label{eqn:jiji}
\end{eqnarray}
Case 2. Assume that an initial state satisfies Eq. (\ref{eqn:tano}). Then Lemma 1 (ii) implies Eq. (\ref{eqn:jiji}). Case 3. Assume that an initial state satisfies Eq. (\ref{eqn:oroti}). Let $\widetilde{\varphi}_1 = ( |\alpha| e^{i\theta_{\alpha}},$ $|\beta| e^{i\theta_{\beta}},$ $|\beta| e^{i(\pi-\theta_{\beta}+2 \theta_{\alpha})}).$ So we have $\widetilde{\varphi}_2 = e^{-i\theta_{\alpha}} \widetilde{\varphi}_1 = (|\alpha|,$ $|\beta| e^{i(\theta_{\beta}-\theta_{\alpha})},$ $|\beta| e^{i(\pi-\theta_{\beta}+\theta_{\alpha})}).$ We put $\theta=\theta_{\beta}-\theta_{\alpha}$, then $\widetilde{\varphi}_2=(|\alpha |, |\beta |e^{i\theta}, |\beta |e^{i(\pi-\theta)})$. Therefore, real and imaginary parts of $\widetilde{\varphi}_2$ become ${\rm Re} (\widetilde{\varphi}_2) = (|\alpha|, |\beta| \cos \theta, -|\beta| \cos \theta)$ and ${\rm Im}(\widetilde{\varphi}_2) = (0, |\beta| \sin\theta, |\beta| \sin \theta),$ respectively. Here if we take ${\rm Re}(\widetilde{\varphi}_2)$ as an initial state of the RCA, then we can see that distribution of $X(n)$ is symmetric by using Lemma 1 (i). Similarly, using Lemma 1 (ii), the initial state ${\rm Im}(\widetilde{\varphi}_2)$ also gives symmetric distribution. In order to confirm $\widetilde{\varphi}_2 ={\rm Re}(\widetilde{\varphi}_2)+i {\rm Im}(\widetilde{\varphi}_2) \in \widetilde{\Phi}_s$, we can use the linearity of Eq. (\ref{eqn:naomi}) and the fact that if the initial state $(\alpha, \beta, \gamma)$ is real, then the solution $\{X_k(n)\}$ is also real. From $\widetilde{\varphi}_2 \in \widetilde{\Phi}_s$, we can get $\widetilde{\varphi}_1 \in \widetilde{\Phi}_s$. Therefore for any $\widetilde{\varphi} \in \widetilde{\Phi}_{\bot}$, we have $| X_k(n)| =|X_{-k}(n)|$ for any $k \in {\bf Z}$ and $n \in {\bf Z}_+$. Then we conclude 
\begin{eqnarray}
\widetilde{\Phi}_{\bot} \subset \widetilde{\Phi}_s
\label{eqn:jijidayo}
\end{eqnarray}
Finally, a direct computation gives
\begin{eqnarray*}
&& m (1) = |\gamma|^2-|\beta|^2, \quad
   m (2) = 2 \cos^2\theta (|\gamma|^2 - |\beta|^2) \\
&& m (3) = {1 \over 2} (3 \cos^22\theta+2 \cos2\theta+1)(|\gamma|^2 - |\beta|^2)
\\
&& \qquad \qquad \qquad 
- {1 \over 2} \sin \theta \> \sin2 \theta \>
\{ \alpha (\overline{\beta}+\overline{\gamma}) + 
\overline{\alpha}(\beta + \gamma) \}
\end{eqnarray*}
where $\displaystyle{m (n) \equiv \sum_{k=-\infty}^{\infty} k |X_k(n)|^2}$.
 The above equations imply that if $\widetilde{\varphi} \in \widetilde{\Phi}_0$ then $\widetilde{\varphi} \in \widetilde{\Phi}_{\bot}$. So we have 
\begin{eqnarray}
\widetilde{\Phi}_0 \subset \widetilde{\Phi}_{\bot}
\label{eqn:jijidane}
\end{eqnarray}
Combining Eqs. (\ref{eqn:tsuika}), (\ref{eqn:jijidayo}) with Eq. (\ref{eqn:jijidane}) implies $\widetilde{\Phi}_{\bot} \subset \widetilde{\Phi}_s \subset \widetilde{\Phi}_0 \subset \widetilde{\Phi}_{\bot}$, so the proof of Theorem 2 is complete.

\section{Conserved quantity for the RCA }
From now on we assume that $0 < \theta < \pi/2$ for simplicity. First we let
\begin{eqnarray*} 
|| X(n) ||^2 = \sum_{k = - \infty} ^{\infty} |X_k (n)|^2
\end{eqnarray*} 
Here we consider a necessary and sufficient condition on the initial state for a conservation law: $|| X(n) || = c \> (n \ge 0)$ with $c \ge 0.$ To do so, we introduce the following two subsets of $\widetilde{\Phi}$:
\begin{eqnarray}
\widetilde{\Phi}_{\ast} (c)= \{  \widetilde{\varphi} \in 
\widetilde{\Phi} : 
&&
|\alpha|^2=c, 
\label{eqn:akiraa}
\\
&&
|\beta|^2+|\gamma|^2=c, 
\label{eqn:akirab}
\\
&&
\beta\overline{\gamma}+\overline{\beta}\gamma=0,
\label{eqn:akirac}
\\
&&
\alpha(\overline{\beta} -\overline{\gamma})+\overline{\alpha}(\beta -\gamma)=2 c \> \cos \theta 
 \quad \}
\label{eqn:akirad}
\end{eqnarray}
and
\[
\widetilde{\Phi}(c)= \left\{ \widetilde{\varphi} \in 
\widetilde{\Phi} : 
|| X(n) ||^2=c \>\> \hbox{for any} \> n \in {\bf Z}_+\right\}
\]
where $\widetilde{\Phi} (c)$ stands for the set of initial states statisfying that $|| X(n) ||$ becomes a conserved quantity. When $c=0$, it is easily shown that $\widetilde{\Phi}(0)=\widetilde{\Phi}_{\ast} (0) = 
\left\{ \widetilde{\varphi} \in \widetilde{\Phi} : \alpha = \beta = \gamma = 0 \right\}.$ So we assume that $c >0.$ Then we have
\begin{thm}
\label{thm:thm3} For any $c>0$, 
\[
\widetilde{\Phi}_{\ast} (c) = \widetilde{\Phi}(c)
\]
\end{thm}
\noindent
The above theorem implies that $\widetilde{\Phi}_{\ast} (c)$ gives a necessary and sufficient condition we want to know. 
\\
Proof. Case 1: $\widetilde{\Phi}(c) \subset \widetilde{\Phi}_{\ast} (c).$ A little algebra reveals
\begin{eqnarray}
&& 
|| X(0) ||^2 = |\alpha|^2,  
\label{eqn:ama}
\\
&&
|| X(1) ||^2 = |\beta|^2+|\gamma|^2, 
\label{eqn:terasu}
\\
&&
|| X(2) ||^2 = |\alpha|^2+
2 \cos^2 \theta (|\beta|^2+|\gamma|^2)-\cos^2 \theta (\beta \overline{\gamma}+\overline{\beta}\gamma)
\nonumber
\\
&&
\qquad \qquad \qquad \qquad
- \cos\theta \> \{ \alpha (\overline{\beta}- \overline{\gamma})+\overline{\alpha} (\beta-\gamma) \},
\label{eqn:oumi}
\\
&&
|| X(3) ||^2 = 2 \cos^2\theta  |\alpha|^2+
\left\{2\cos^4\theta + (1-2 \cos^2\theta)^2\right\} (|\beta|^2+|\gamma|^2)
\nonumber
\\
&&
\qquad \qquad \qquad
+2 \cos^2\theta (1-2 \cos^2\theta) (\beta\overline{\gamma}+\overline{\beta}\gamma)
\nonumber
\\
&&
\qquad \qquad \qquad
+\cos\theta (1-3 \cos^2\theta) 
\{ \alpha (\overline{\beta}-\overline{\gamma})+
\overline{\alpha}(\beta-\gamma)\}
\label{eqn:kami}
\end{eqnarray}
From Eqs. (\ref{eqn:ama}) - (\ref{eqn:oumi}) and $0 < \theta < \pi/2,$ we have
\begin{eqnarray} 
\alpha (\overline{\beta}-\overline{\gamma})+
\overline{\alpha}(\beta-\gamma) = \cos \theta \{  2c -(\beta\overline{\gamma}+\overline{\beta}\gamma) \} 
\label{eqn:taishi}
\end{eqnarray}
Combining Eq. (\ref{eqn:taishi}) with Eqs. (\ref{eqn:ama}), (\ref{eqn:terasu}), (\ref{eqn:kami}) implies $\sin \theta \cos \theta ( \beta\overline{\gamma}+\overline{\beta}\gamma )= 0,$ so we have $\beta\overline{\gamma}+\overline{\beta}\gamma = 0$, since $0 < \theta < \pi/2.$ Then Eq. (\ref{eqn:taishi}) becomes $\alpha (\overline{\beta}-\overline{\gamma}) + \overline{\alpha}(\beta-\gamma) = 2 c \> \cos \theta.$ So we have the desired result.
\par\noindent
Case 2: $\widetilde{\Phi}_{\ast}(c) \subset \widetilde{\Phi}(c).$ Define a spacial discrete Fourier transformation of $X_k(n)$ by 
\begin{eqnarray*} 
\widetilde{X}_n (\xi) = \sum_{k = - \infty} ^{\infty} e^{i \xi k} X_k (n)
\end{eqnarray*} 
where $\xi \in {\bf R}.$ From Eq. (\ref{eqn:naomi}), we have 
\begin{eqnarray*} 
\widetilde{X}_{n+2} (\xi) = \cos \theta (e^{-i \xi} - e^{i \xi}) 
\widetilde{X}_{n+1} (\xi) + \widetilde{X}_{n} (\xi)
\end{eqnarray*} 
where $\widetilde{X}_{0} (\xi) = \alpha$ and $\widetilde{X}_{1} (\xi) = e^{-i \xi} \beta + e^{i \xi} \gamma.$ We define $\varphi = \varphi (\xi) \in {\bf R}$ by $\lambda_+ (\xi)=e^{i \varphi}$, that is, $\cos \varphi = \sqrt{1 - \cos^2 \theta \sin^2 \xi}$ and $\sin \varphi =- \cos \theta \sin \xi.$ So $\lambda_- (\xi)= - e^{- i \varphi}$. Then we obatin
\begin{eqnarray*} 
\widetilde{X}_{n} (\xi) = A(\xi) \lambda_+ ^n (\xi) + B(\xi) \lambda_- ^n (\xi)
\end{eqnarray*} 
where $\lambda_{\pm} (\xi) = - i \cos \theta \sin \xi \pm \sqrt{1 - \cos^2 \theta \sin^2 \xi}$ and 
\begin{eqnarray*} 
A(\xi) = {\alpha e^{-i \varphi} + \widetilde{X}_{1} (\xi)
\over e^{i \varphi}+e^{- i \varphi}}, \quad
B(\xi) = {\alpha e^{i \varphi} - \widetilde{X}_{1} (\xi)
\over e^{i \varphi}+e^{- i \varphi}}
\end{eqnarray*} 
Define 
\begin{eqnarray*} 
\braa f | g \kett = {1 \over 2 \pi} \int_0 ^{2 \pi} \overline{f(\xi)} g(\xi) d \xi\end{eqnarray*} 
In particular, we put $|| f ||_{\ast} ^2 = \braa f | f \kett.$ Then
\begin{eqnarray*} 
\sum_{k = - \infty} ^{\infty} |X_k (n)|^2 = 
{1 \over 2 \pi} \int_0 ^{2 \pi} | \widetilde{X}_{n} (\xi)|^2 d \xi
\end{eqnarray*} 
implies $|| X(n) || = || \widetilde{X}_{n} ||_{\ast}.$ In this setting, we have the following result: for any $n \in {\bf Z}_+$, 
\begin{eqnarray*} 
|| X(n) ||^2 = || A ||_{\ast} ^2 + ||B||_{\ast}^2 +(-1)^n 
\{ \braa e^{i n \varphi} A| e^{- i n \varphi} B \kett 
+ \braa e^{- i n \varphi} B| e^{i n \varphi} A \kett \}
\end{eqnarray*} 
A direct computation gives
\begin{eqnarray*} 
&& 
||A||_{\ast} ^2=||B||_{\ast} ^2 
\\
&&
={1 \over 4 \sin \theta} 
\left[ 
(| \alpha |^2 + | \beta |^2 + | \gamma |^2 )
- \{ \alpha (\overline{\beta} - \overline{\gamma})
+ \overline{\alpha}(\beta - \gamma) \} 
\left({ 1 - \sin \theta \over \cos \theta } \right) 
\right.
\\
&&
\qquad \qquad \qquad \qquad \qquad \qquad \qquad \qquad \qquad \qquad 
\left.
- (\beta \overline{\gamma}+ \overline{\beta} \gamma) 
\left( { 1 - \sin \theta \over \cos \theta } \right)^2 
\right] 
\end{eqnarray*} 
Moreover we have
\begin{eqnarray*} 
&&
4 \pi \{ \braa e^{i n \varphi} A| e^{- i n \varphi} B \kett 
+ \braa e^{- i n \varphi} B| e^{i n \varphi} A \kett \}
\\
&&
=
| \alpha |^2 
\int_0 ^{2 \pi} {\cos (2(n-1)\varphi) \over \cos^2 \varphi} \> d \xi
\\
&&
+( \overline{\alpha} \gamma - \alpha \overline{\beta}) i
\left\{
\int_0 ^{2 \pi} {\cos \xi \sin ((2n-1) \varphi) \over \cos^2 \varphi} \> d \xi
+i 
\int_0 ^{2 \pi} {\sin \xi \sin ((2n-1) \varphi) \over \cos^2 \varphi} \> d \xi
\right\}
\\
&&
+( \overline{\alpha} \beta - \alpha \overline{\gamma}) i
\left\{
\int_0 ^{2 \pi} {\cos \xi \sin ((2n-1) \varphi) \over \cos^2 \varphi} \> d \xi
-i 
\int_0 ^{2 \pi} {\sin \xi \sin ((2n-1) \varphi) \over \cos^2 \varphi} \> d \xi
\right\}
\\
&&
-(| \beta |^2 +| \gamma |^2)
\int_0 ^{2 \pi} {\cos (2n \varphi) \over \cos^2 \varphi} \> d \xi
\\
&&
- \beta \overline{\gamma} 
\left\{
\int_0 ^{2 \pi} {\cos (2 \xi) \cos (2n \varphi) \over \cos^2 \varphi} \> d \xi
-i 
\int_0 ^{2 \pi} {\sin (2 \xi) \cos (2n \varphi) \over \cos^2 \varphi} \> d \xi
\right\}
\\
&&
- \overline{\beta} \gamma
\left\{
\int_0 ^{2 \pi} {\cos (2 \xi) \cos (2n \varphi) \over \cos^2 \varphi} \> d \xi
+i 
\int_0 ^{2 \pi} {\sin (2 \xi) \cos (2n \varphi) \over \cos^2 \varphi} \> d \xi
\right\}
\end{eqnarray*}
for any $n \in {\bf Z}_+$. Using a change of variable in the integration and noting
\[
\int_0^{2\pi}{\cos\xi\sin((2n-1)\varphi) \over \cos^2\varphi}d\xi
= \int_0^{2\pi}{\sin(2\xi)\cos(2n\varphi) \over \cos^2\varphi}d\xi=0
\]
we have
\begin{eqnarray} 
|| X(n) ||^2
&=&
{1 \over 2 \sin \theta} 
\Biggl[ (| \alpha |^2 + | \beta |^2 + | \gamma |^2 )
\nonumber
\\
&&
\qquad \qquad 
- \{ \alpha (\overline{\beta} - \overline{\gamma})+
\overline{\alpha}(\beta - \gamma) \} 
\left({ 1 - \sin \theta \over \cos \theta } \right)
\nonumber
\\
&&
\qquad \qquad \qquad \qquad \qquad \qquad 
- (\beta \overline{\gamma}+ \overline{\beta} \gamma) 
\left( { 1 - \sin \theta \over \cos \theta } \right)^2 
\Biggr] 
\nonumber
\\
&&
+
{(-1)^n \over \pi} \left[
| \alpha |^2 
\int_{\theta - \pi/2} ^{0} 
{\cos (2(n-1)x) \over \cos x \sqrt{\cos ^2 x - \sin ^2 \theta}} \> dx
\right.
\nonumber
\\
&&
- \{ \alpha (\overline{\beta} - \overline{\gamma})+
\overline{\alpha}(\beta - \gamma) \} 
{1 \over \cos \theta} 
\int_{\theta - \pi/2} ^{0} 
{\sin x \sin ((2n-1)x) \over \cos x \sqrt{\cos ^2 x - \sin ^2 \theta}} \> dx
\nonumber
\\
&&
-(| \beta |^2 +| \gamma |^2)
\int_{\theta - \pi/2} ^{0} 
{\cos (2nx) \over \cos x \sqrt{\cos ^2 x - \sin ^2 \theta}} \> dx
\nonumber
\\
&&
-(\beta \overline{\gamma}+ \overline{\beta} \gamma) 
\left\{ 
{2 \over \cos^2 \theta} 
\int_{\theta - \pi/2} ^{0} 
{\cos (2nx) \sqrt{\cos ^2 x - \sin ^2 \theta} \over \cos x} \> dx
\right.
\nonumber
\\
&&
\left.
\left.
\qquad \qquad \qquad \qquad 
- \int_{\theta - \pi/2} ^{0} 
{\cos (2nx) \over \cos x \sqrt{\cos ^2 x - \sin ^2 \theta}} \> dx
\right\}
\right]
\label{eqn:norio}
\end{eqnarray}
for any $n \in {\bf Z}_+$. The condition ``$|\alpha|^2 = |\beta|^2+|\gamma|^2=c, \> \alpha (\overline{\beta} - \overline{\gamma})+\overline{\alpha}(\beta - \gamma)  = 2 c \> \cos \theta, \> \beta \overline{\gamma}+ \overline{\beta} \gamma =0$'' in $\widetilde{\Phi}_{\ast} (c)$ gives
\begin{eqnarray*} 
|| X(n) ||^2
= c + {(-1)^n c \over \pi}
\int_{\theta - \pi/2} ^{0} 
{h_n (x) \over \cos x \sqrt{\cos ^2 x - \sin ^2 \theta}} 
\> dx
\end{eqnarray*}
where $h_n (x) = \cos (2(n-1)x) - 2 \sin x \sin((2n-1)x) - \cos(2nx).$ On the other hand, a little algebra reveals that $h_n (x) = 0$ for any $n \in {\bf Z}_+.$ Therefore we conclude that $|| X(n) ||^2 = c \> (n=0,1,2,  \ldots),$ that is, $\widetilde{\Phi}_{\ast} (c)\subset \widetilde{\Phi}(c).$ The proof of Theorem 3 is complete.
\par
By using Eq. (\ref{eqn:norio}) and the Riemann-Lebesgue lemma (see page 462 in Durrett \cite{Durrett}, for example), we have
\begin{eqnarray} 
&&
\lim_{n \to \infty} || X(n) ||^2
=
{1 \over 2 \sin \theta} 
\Biggl[ (| \alpha |^2 + | \beta |^2 + | \gamma |^2 )
\nonumber
\\
&&    
- \{ \alpha (\overline{\beta} - \overline{\gamma})+
\overline{\alpha}(\beta - \gamma) \} 
\left({ 1 - \sin \theta \over \cos \theta } \right)
- (\beta \overline{\gamma}+ \overline{\beta} \gamma) \left( { 1 - \sin \theta \over \cos \theta } \right)^2 
\Biggr] 
\label{eqn:masako}
\end{eqnarray} 
The above equation will be used in next section. As a corollary to Theorem 3, we can show that any RCA with symmetric distribution does not have our conserved quantity:
\begin{cor} For any $c>0$, 
\[
\widetilde{\Phi}_s \cap \widetilde{\Phi}(c)= \phi
\]
\end{cor}
\par\noindent
Proof. We consider the following three cases. Case 1. When $\beta + \gamma =0$, from $|\beta|^2+|\gamma|^2=c$, we have $\beta = \sqrt{c/2} \> e^{i \theta_{\beta}}$ and $\gamma = - \sqrt{c/2} \> e^{i \theta_{\beta}}.$ However the last equations and $\beta \overline{\gamma} + \overline{\beta} \gamma = 0$ contradict each other, since $\beta \overline{\gamma} + \overline{\beta} \gamma = - c \> (<0)$. Case 2. ``$|\beta|= |\gamma|(>0),$ and $\alpha =0$'' and ``$|\alpha|^2 = c \> (>0)$'' contradict each other. Case 3. We assume that ``$ |\beta|= |\gamma|(>0), \alpha \not= 0, \> \hbox{and} \>\> \theta_{\beta}+\theta_{\gamma}-2\theta_{\alpha} =\pi \>\> ( \hbox{mod} \> 2\pi)$''. From now on we omit ``$\hbox{mod} \> 2\pi$''. Then $\alpha(\overline{\beta} -\overline{\gamma})+\overline{\alpha}(\beta -\gamma)=2 c \> \cos \theta$ can be rewritten as 
\begin{eqnarray}  
|\alpha| \{ |\beta| \cos (\theta_{\alpha} - \theta_{\beta}) - 
|\gamma| \cos (\theta_{\alpha} - \theta_{\gamma}) \} = c \> \cos \theta  
\label{eqn:masae}
\end{eqnarray}  
On the other hand, $|\alpha|=\sqrt{c}, \> |\beta|=|\gamma|=\sqrt{c/2}$. So Eq. (\ref{eqn:masae}) becomes
\begin{eqnarray}  
\cos (\theta_{\alpha} - \theta_{\beta}) - 
\cos (\theta_{\alpha} - \theta_{\gamma}) = \sqrt{2} \> \cos \theta  
\label{eqn:mannosuke}
\end{eqnarray}  
Combining Eq. (\ref{eqn:mannosuke}) with $\theta_{\beta}+\theta_{\gamma}-2\theta_{\alpha} =\pi$ implies
\begin{eqnarray}  
\sin \left( (\theta_{\beta}- \theta_{\gamma})/2 \right)  
= - \cos \theta /\sqrt{2}  
\label{eqn:himiko}
\end{eqnarray} 
We see that $\beta \overline{\gamma} + \overline{\beta} \gamma = 0$ gives $\theta_{\beta}- \theta_{\gamma} = \pm \pi/2$. So Eq. (\ref{eqn:himiko}) becomes $\cos \theta = \pm 1$. Then we have a contradiction, since we assumed that $0 < \theta < \pi/2.$
 So the proof of Corollary 4 is complete.

\section{The QW case}
In the last section we return to the QW given by $H(\theta)$. So we put initial qubit state $\varphi = {}^t [\alpha_l, \alpha_r]$ with $|\alpha_l|^2+|\alpha_r|^2=1$ at the origin for the QW. Here we apply our result (Theorem 3) on conservation property of the RCA to the QW. First we consider the left chirality case, that is, $\alpha = \alpha_l, \> \beta = \cos \theta \> \alpha_l + \sin \theta \> \alpha_r$ and $\gamma = 0$. Then Eq. (\ref{eqn:akirac}) holds, since $\gamma =0.$ Eqs. (\ref{eqn:akirab}) and (\ref{eqn:akirad}) can be rewritten as 
\begin{eqnarray}
&& 
\cos ^2 \theta |\alpha_l|^2 + \sin ^2 \theta |\alpha_r|^2 + \cos \theta \sin \theta (\alpha_l \overline{\alpha_r} + \overline{\alpha_l} \alpha_r ) = c
\label{eqn:akirabb}
\\
&&
2 \cos \theta |\alpha_l|^2 + \sin \theta (\alpha_l \overline{\alpha_r} + \overline{\alpha_l} \alpha_r ) = 2 c \cos \theta 
\label{eqn:akiradd}
\end{eqnarray} 
respectively. By using $\sin \theta \not=0$ (since $0 < \theta < \pi/2$) and Eqs. (\ref{eqn:akiraa}) and (\ref{eqn:akiradd}), we have 
\begin{eqnarray} 
\alpha_l \overline{\alpha_r} + \overline{\alpha_l} \alpha_r =0
\label{eqn:akiraee}
\end{eqnarray} 
Combining Eq. (\ref{eqn:akirabb}) with Eq. (\ref{eqn:akiraee}) gives 
\begin{eqnarray} 
\cos^2 \theta |\alpha_l|^2 + \sin^2 \theta |\alpha_r|^2 =c
\label{eqn:hime}
\end{eqnarray} 
On the other hand, it follows from $|\alpha_l|^2+|\alpha_r|^2=1$ and Eq. (\ref{eqn:akiraa}) that $|\alpha_r|^2 = 1-c.$ By Eq. (\ref{eqn:hime}) and the last equation, we get 
\begin{eqnarray} 
c=1/2
\label{eqn:kaguyahime}
\end{eqnarray}  
Moreover $|\alpha_l|=|\alpha_r|=1/\sqrt{2}$ can be derived from Eq. (\ref{eqn:kaguyahime}). So combining this result and Eq. (\ref{eqn:akiraee}), we conclude that the initial state of the QW given by the RCA having our conserved quantity with $c=1/2$ can be determined as 
\begin{eqnarray} 
\varphi =
\left[
\begin{array}{cc}
\alpha_l \\
\alpha_r  
\end{array}
\right]
= \pm {e^{i \xi} \over \sqrt{2}}
\left[
\begin{array}{cc}
1 \\
i  
\end{array}
\right]
\label{eqn:akiraff}
\end{eqnarray} 
where $\xi \in {\bf R}.$ Similarly, we consider the right chirality case, that is, $\alpha = \alpha_r, \> \beta = 0,$ and $\gamma = \sin \theta \> \alpha_l - \cos \theta \> \alpha_r$. In this case also, a similar computation gives the same conclusions, i.e., Eqs. (\ref{eqn:kaguyahime}) and (\ref{eqn:akiraff}). 

On the other hand if we choose Eq. (\ref{eqn:akiraff}) as an initial qubit state for the QW defined by $H(\theta)$, then the probability distribution becomes symmetric (see Refs. \cite{KonA,KonB,KonC}).  Furthermore our result gives an additional information on the conserved quantity:
\[
|| \Psi^L (n) ||^2 =|| \Psi^R (n) ||^2= {1 \over 2}
\] 
for any $n \ge 0.$ In particular, when $\theta = \pi/4$ (the Hadamard walk), if $\varphi = {}^t[1/\sqrt{2},i/\sqrt{2}]$, then $|| \Psi^L (n) ||^2 =|| \Psi^R (n) ||^2 = 1/2$ for any $n \ge 0.$ However Corollary 4 implies that the conserved quantity (in our case $c=1/2$) is not compatible with the symmetry of the distribution. This gives an interesting result on the QW: any symmetric probability distribution of the QW can not be written as the sum of the symmetric distribution for each chirality of the QW. 

Finally we consider a general case. The unitary evolution of the QW implies $
|| \Psi^L (n) ||^2 +|| \Psi^R (n) ||^2=1$ for any time step $n \in {\bf Z}_+.$ Furthermore Eq. (\ref{eqn:masako}) gives
\begin{eqnarray*} 
\lim_{n \to \infty} || \Psi^L (n) ||^2
&=&
{1 \over 2 \sin \theta} 
\left[ (1+ \cos^2 \theta) | \alpha_l |^2 + \sin ^2 \theta | \alpha_r |^2 
\right.
\\
&&
\qquad  
+ \sin \theta \cos \theta 
(\alpha_l \overline{\alpha_r} + \overline{\alpha_l} \alpha_r )
\\
&& 
\qquad 
\left.
- \{ 2 \cos \theta |\alpha_l|^2 
+ \sin \theta (\alpha_l \overline{\alpha_r} + \overline{\alpha_l} \alpha_r ) \}
\left( { 1 - \sin \theta \over \cos \theta } \right) \right] 
\\
\lim_{n \to \infty} || \Psi^R (n) ||^2
&=&
{1 \over 2 \sin \theta} 
\left[ (1+ \cos^2 \theta) | \alpha_r |^2 + \sin ^2 \theta | \alpha_l |^2 
\right.
\\
&&
\qquad 
- \sin \theta \cos \theta 
(\alpha_l \overline{\alpha_r} + \overline{\alpha_l} \alpha_r )
\\
&& 
\qquad  
\left.
- \{ 2 \cos \theta |\alpha_r|^2 
- \sin \theta (\alpha_l \overline{\alpha_r} +  \overline{\alpha_l} \alpha_r) \}
\left( { 1 - \sin \theta \over \cos \theta } \right) \right] 
\end{eqnarray*} 
For example, if we choose $\theta \in (0, \pi/2)$ and $\varphi = {}^t[1,0]$ (asymmetric case), then
\[
\lim_{n \to \infty} || \Psi^L (n) ||^2 = 1 - {\sin \theta \over 2} \in \left({1 \over 2},1 \right), 
\quad \lim_{n \to \infty} || \Psi^R (n) ||^2 = {\sin \theta \over 2} \in \left(0, {1 \over 2} \right)
\]
\par
\
\par\noindent
{\bf Acknowledgment.}  
\par\noindent
This work is partially financed by the Grant-in-Aid for Scientific Research (B) (No.12440024) of Japan Society of the Promotion of Science.



\end{document}